\documentclass[aps, pra, twocolumn,showpacs,preprintnumbers]{revtex4}
    \usepackage{graphicx,amsmath,amssymb,amsthm,bbm}
    
\def\refeqn#1{Eq.\ (\ref{Equation::#1})}

\def\refsec#1{Section \ref{Section::#1}}

\def\refapx#1{Appendix \ref{Appendix::#1}}

\def\reffig#1{Figure \ref{Figure::#1}}
\def\reffigs#1#2{Figures \ref{Figure::#1} and \ref{Figure::#2}}

\begin{document}
    \title{Deterministic Dicke state preparation with continuous measurement and control}
    \author{John K. Stockton}
    \homepage{http://minty.caltech.edu/Ensemble}
    \email{jks@caltech.edu}
    \author{Ramon van Handel}
    \author{Hideo Mabuchi}
    \affiliation{Norman Bridge Laboratory of Physics, M.C.
    12-33, California Institute of Technology, Pasadena CA 91125}

\date{\today}

\begin{abstract}

We characterize the long-time projective behavior of the stochastic master equation describing a continuous, collective spin measurement of an atomic ensemble both analytically and numerically.  By adding state based feedback, we show that it is possible to prepare highly entangled Dicke states deterministically.

\end{abstract}


\maketitle

\section{\label{Section::Introduction}INTRODUCTION}

It has long been recognized that measurement can be used as a {\em non-deterministic} means of preparing quantum states that are otherwise difficult to obtain. With projective measurements that are truly discrete in time, the only way an experimentalist can direct the outcome of the measurement is by preparing the initial state to make the desired result most probable.  Generally, it is impossible to make this probability equal to one, as the measurement will, with some non-zero probability, result in other undesirable states.  If the experimentalist can afford to be patient, then accepting a low efficiency is not a problem, but this is not always the case.  In recent years, a theory of continuous quantum measurement has been developed that fundamentally changes the nature of state preparation via measurement \cite{Wiseman1996}.  When a measurement and the corresponding acquisition of information are sufficiently gradual, there exists a window of opportunity for the experimentalist to affect the outcome of the measurement by using feedback control \cite{Wiseman1994}.  In this paper, we demonstrate that it is possible to deterministically prepare highly entangled Dicke states \cite{Dicke1954, Mandel1995} of an atomic spin ensemble by adding state based feedback to a continuous projective measurement.

It has been shown that models of quantum state reduction exist that exhibit the usual rules of projective measurement except the state reduction occurs in a continuous, stochastic manner \cite{Hughston2001}.  These models are not without physical relevance as they are the same as those derived to describe the conditional evolution of atomic spin states under  continuous quantum non-demolition (QND) measurement \cite{Thomsen2002a, Thomsen2002b}.  By measuring the collective angular momentum operator, $J_z$, of an initially polarized coherent spin state via the phase shift of an off-resonant probe beam it is possible to produce conditional spin squeezed states \cite{Kuzmich2000, Geremia2004a} which are of considerable interest for applications in quantum information processing and precision metrology \cite{Wineland1994, Polzik2003}.

In these models, the reduction in variance that initially leads to conditional spin squeezing is the pre-cursor of the projection onto a random eigenstate of $J_z$ at longer times.  \reffig{1} demonstrates the projection process for a single numerically simulated measurement trajectory \footnote{All numerical simulations shown were performed using the parameters $\{N=10,\,M=1,\,T=5,\,dt=0.001\}$.  The stochastic integrator used the norm-preserving, non-linear SSE of \refeqn{SSE} and a weak second-order derivative-free predictor-corrector structure as can be found in \cite{Kloeden2003}.}. Like spin squeezed states, these Dicke states offer potential for quantum information applications because of their unique entanglement properties \cite{Stockton2003a}. Although the experimental difficulties in obtaining these states via QND measurement or other experimental methods \cite{Duan2003, Molmer2003,Unanyan2002} are considerable, the details of the continuous projective process that leads to them are of fundamental interest.

Whenever the measurement is sufficiently slow, an experimentalist may steer the result by feeding back the measurement results in real time to a Hamiltonian parameter. Indeed, the measurement process, as a state preparation process, can be made deterministic with the use of feedback control.  Recently, we have demonstrated this concept by modulating a compensation magnetic field with the measurement record to deterministically prepare spin squeezed states \cite{Geremia2004a} as proposed in \cite{Thomsen2002a, Thomsen2002b}.  This is just one example of the growing confluence of quantum measurement with classical estimation and control theory \cite{Doherty2000, Belavkin1999}.  Other applications of quantum feedback include parameter estimation, metrology, and quantum error correction \cite{Geremia2003a, Stockton2003b, Andre2004, Armen2002, Ahn2002}.

\begin{figure*}
\includegraphics[width=7in]{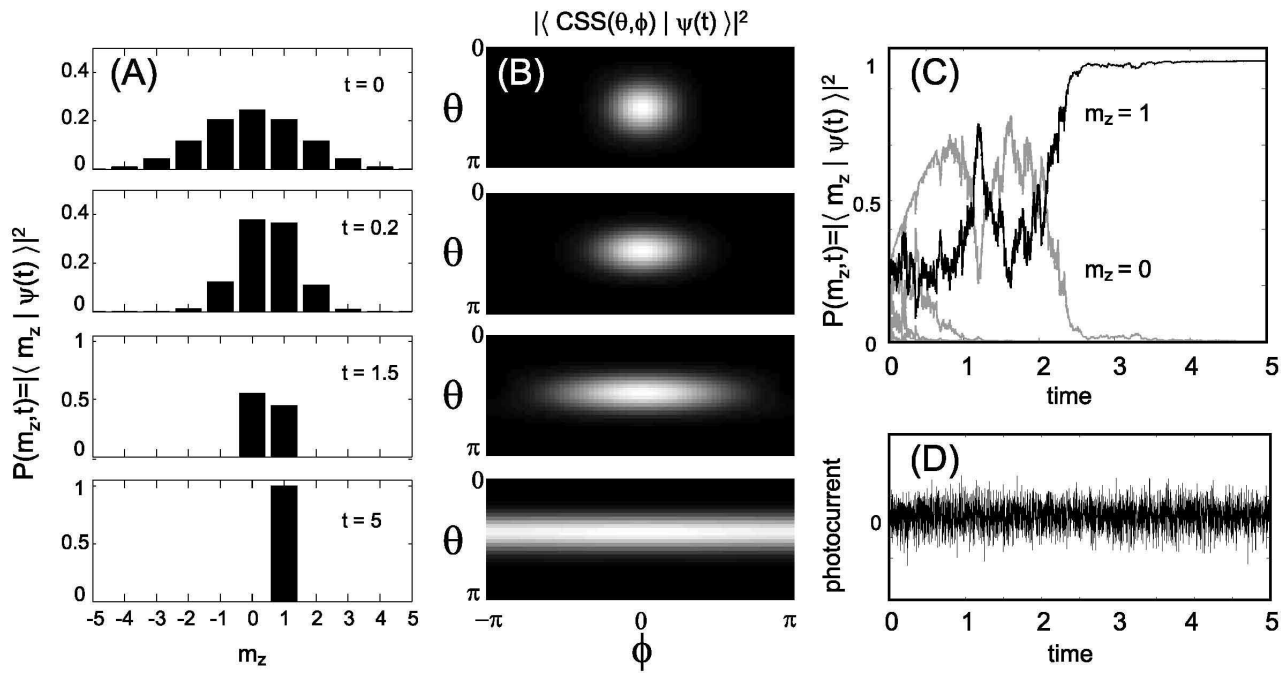} \caption{ \label{Figure::1} The results of a single numerical simulation [36] of the SSE, \refeqn{SSE}, with $M=1$, $\eta=1$, and $N=10$ spins initially aligned along the $x$-axis. (A) In a quantization axis perpendicular to the polarization, the level distribution of a coherent spin state (CSS) is Gaussian for large $N$.  Under conditional measurement the state evolves at short times into a spin squeezed state and, eventually, into a random eigenstate of $J_z$.  (B) A map of the state's angular distribution on the Bloch sphere in spherical coordinates. The uncertainty in the transverse direction to the measurement axis grows until there is no information about the perpendicular component direction.  (C) At long times, the population is at most divided among two levels that compete to be the final winner, which in this case appears to be $m=1$. (D) All of the state information is obtained by properly filtering the noisy photocurrent.} 
\end{figure*}

In this paper, we focus on the long time limit of the QND measurement and feedback process.  Just as spin squeezed states can be deterministically prepared at short times, we numerically demonstrate that individual Dicke states can be deterministically prepared at long times with the use of state based feedback \cite{Doherty1999}. While our proposed feedback laws are non-optimal, they demonstrate the adequacy of intuitive controllers with finite gain for directing the diffusion of the quantum state towards desirable regions of Hilbert space with unity efficiency.  This is in contrast to other proposed schemes using measurement to prepare Dicke states probabilistically \cite{Duan2003, Molmer2003}.  A more systematic approach utilizing stochastic notions of stability and convergence in the continuous measurement and control of a single spin is presented in reference \cite{vanHandel2004}.

This paper is organized as follows.  In \refsec{SME}, we introduce the stochastic master equation which represents the rule for updating the system state in real time via the incoming measurement record.  Here we discuss the various representations of the dynamics in both the short and long time limits.  \refsec{OL} describes the probabilistic preparation of Dicke states via observation alone. The numerical demonstration of the open loop projection process reveals statistical features that clarify the details of the projection. Feedback is added to the procedure in \refsec{CL}, where we show that state based control allows one to prepare the same Dicke state deterministically on every measurement. Finally, in \refsec{Conclusion}, we discuss future directions and imminent challenges regarding quantum state preparation via measurement and control.

\section{\label{Section::SME}REPRESENTATIONS OF THE CONDITIONAL EVOLUTION}

The physical system we will consider is an ensemble of $N$ spin-1/2 particles contained within a cavity and interacting with a far off resonant single field mode.  We will denote the conditional state of the spin ensemble as $\rho(t)$ and the homodyne measurement record of the output as $y(t)$.  The stochastic master equation (SME) describing the conditional evolution is \cite{Thomsen2002a,Thomsen2002b}
\begin{eqnarray}
d\rho(t)=-i[H(t),\rho(t)]dt+\mathcal{D}[\sqrt{M}J_z]\rho(t)dt\nonumber\\
+\sqrt{\eta}\mathcal{H}[\sqrt{M}J_z]\rho(t)\left(2 \sqrt{M\eta}[y(t) dt-\langle J_z\rangle dt]\right)\label{Equation::SME}
\end{eqnarray}
where $H(t)=\gamma J_y b(t)$ is the control Hamiltonian that we will allow ourselves (without feedback $b(t)=0$), $\gamma$ is the gyromagnetic ratio, $M$ is the probe parameter dependent measurement rate, and
\begin{eqnarray}
\mathcal{D}[c]\rho&\equiv& c \rho c^{\dagger}-(c^{\dagger}c\rho+\rho c^{\dagger}c)/2\\
\mathcal{H}[c]\rho&\equiv&c\rho+\rho c^{\dagger}-\textrm{Tr}[(c+c^{\dagger})\rho]\rho
\end{eqnarray}
The (scaled) difference photocurrent is represented as
\begin{equation}
y(t) dt =\langle J_z\rangle (t) dt + dW(t)/2 \sqrt{M\eta}\label{Equation::Photocurrent}
\end{equation}
The stochastic quantity $dW(t) \equiv 2 \sqrt{M\eta}(y(t) dt-\langle J_z\rangle(t) dt)$ is a Wiener increment and $dW(t)/dt$ is a Gaussian white noise that can be identified with the shot-noise of the homodyne local oscillator. (See \cite{Oksendal1998, Gardiner2002} for an introduction to stochastic differential equations (SDE).) The sensitivity of the photodetection per $\sqrt{\textrm{Hz}}$ is represented by $1/2\sqrt{M\eta}$, where the quantity $\eta \in [0,1]$ represents the quantum efficiency of the detection.  If $\eta = 0$, we are essentially ignoring the measurement result and the conditional SME becomes a deterministic unconditional master equation.  If $\eta=1$, the detectors are maximally efficient.  In this latter case, the conditioned state will remain pure for the entire measurement, thus we can use a state vector description, and the SME can be replaced with a stochastic Schrodinger equation (SSE)
\begin{eqnarray}
d |\psi (t) \rangle&=&(-i H(t) -M(J_z-\langle J_z \rangle (t))^2/2)|\psi (t) \rangle dt \nonumber \\
&&+\sqrt{M}(J_z-\langle J_z \rangle (t))|\psi (t) \rangle dW(t) \label{Equation::SSE}
\end{eqnarray}
This SSE was considered in \cite{Hughston2001} where the motivation was more abstract and less concerned with the experimental filtering perspective presented here. We emphasize that the SME/SSE is physically derived and is an explicit function of a measured photocurrent variable $y(t)$, through which the randomness enters. The states are considered as states of knowledge and, in practice, an experimentalist updates the description of the system, $\rho(t)$ (\reffig{1}A-C), as the measurement results, $y(t)$ (\reffig{1}D), arrive in time.  

Even though the light field is dispersively coupled to each atom, there will still be some absorption of light and hence decoherence for which the above master equation does not account.  If the cavity coupling is strong enough, and extraneous noise sources are sufficiently reduced, the measurement rate will be faster than this decoherence and we can consider the long time limit of the master equation \cite{Thomsen2002a, Thomsen2002b}.  For free space measurements \cite{Jessen2003, Deutsch2003}, e.g. free space Faraday rotation or homodyne detection of one polarization component, the above SME will be approximately valid in the short time limit only.

\subsection{\label{Section::Dicke}Hilbert space, coherent spin states, and Dicke states}

Under certain idealizations, we can considerably reduce the size of the Hilbert space needed to describe the conditionally measured ensemble. Throughout this paper, the initial state $\rho(0)$ will be made equal to a coherent spin state (CSS) polarized along an arbitrary direction \cite{Mandel1995}.  For example, a CSS pointing along the $z$-axis is denoted $|\uparrow_1\uparrow_2\cdots\uparrow_N\rangle_z$ and all others can be prepared by rotating this state with the angular momentum operators $J_i$, with $i\in\{x,y,z\}$. A CSS, typically obtained via a dissipative optical pumping process, is an eigenstate of $\mathbf{J}^2$ with maximal eigenvalue $J(J+1)$, where $J=N/2$. Because the SME works under the QND approximation of negligible absorption (i.e. the  large detuning dispersive limit), no angular momentum will be exchanged between the probe beam and the ensemble.  The measurement and possible field rotations represent the only allowed dynamics, thus the state will maintain maximal $\langle\mathbf{J}^2 \rangle$ over the course of the measurement.

The Dicke states are defined \cite{Mandel1995} as the states $|l,m\rangle$ that are simultaneous eigenstates of both $\mathbf{J}^2$ and $J_z$
\begin{eqnarray}
J_z |l,m\rangle &=& m |l,m\rangle\\
\mathbf{J}^2 |l,m\rangle &=& l(l+1) |l,m\rangle
\end{eqnarray}
where
\begin{equation}
|m|\leq  l  \leq J=N/2
\end{equation}
Under the above approximations, we can neglect any state with $l\neq J$.  We then shorten the labelling of our complete basis from $|J,m\rangle$ to $|m\rangle$ so that
\begin{eqnarray}
J_z |m\rangle &=& m |m\rangle\\
\mathbf{J}^2 |m\rangle &=& J(J+1) |m\rangle
\end{eqnarray}
where $m \in \{-N/2,-N/2+1, ..., N/2-1, N/2\}$.

When the physical evolution is such that the $|m\rangle$ states remain complete, we can limit ourselves to a density matrix of size $(N+1)\times(N+1)$ rather than the full size $2^N\times 2^N$. This reduced space is referred to as the symmetric subspace, as its states are invariant to particle exchange \cite{Stockton2002, Wiseman2003}. For the case of two spins, the symmetric subspace contains the triplet states, but not the singlet.  States contained within the symmetric subspace can be described as a pseudo-spin of size $J=N/2$.

In the z basis, the extremal values of $m$, $\pm N/2$, are simply the coherent spin states pointing along the $z$-axis
\begin{eqnarray}
|m=+N/2\rangle &=& |\uparrow_1\uparrow_2\cdots\uparrow_N\rangle\\
|m=-N/2\rangle &=& |\downarrow_1\downarrow_2\cdots\downarrow_N\rangle
\end{eqnarray}
In terms of the constituent spins, these states are obviously unentangled. In contrast, consider the state with $m=0$ (for $N$ even)
\begin{equation}
|m=0\rangle = C\,\Sigma_{i} \textrm{P}_i(|\uparrow_1\cdots\uparrow_{N/2}\downarrow_{N/2+1}\cdots\downarrow_N\rangle)
\end{equation}
where the $\textrm{P}_i$ represent all permutations of the spins and $C$ is a normalization constant.  This state is highly entangled in a way that is robust to particle loss \cite{Stockton2003a}.  Even though the expectation values $\langle J_i \rangle$ vanish for this state, it still has maximal $\mathbf{J}^2$ eigenvalue.  Loosely, this state represents a state of knowledge where the length of the spin vector is known and the z component is known to be zero,  but the direction of the spin vector in the $x-y$ plane is completely indeterminate. Similarly, the entangled states with $0<|m|<N/2$ can be imagined as living on cones aligned along the $z$-axis with projection $m$.  The loss of pointing angle information from the measurement process is diagrammed in \reffig{1}B.

Along with their unique entanglement and uncertainty properties, Dicke states are also of interest for the important role they play in descriptions of collective radiation processes \cite{Mandel1995} and for their potential role in quantum information processing tasks \cite{Cabello2002, Molmer2003, Duan2003}.

\subsection{\label{Section::ShortTime}Short time limit}

Even when working within the symmetric subspace, for a large number of spins the size of $\rho(t)$ may be too unwieldy for computational efficiency.  Because it is often desirable to update our state description in real-time (e.g., for optimal feedback procedures), finding simple but sufficient descriptors is of considerable importance.

We can derive a reduced model by employing a moment expansion for the observable of interest. Extracting the conditional expectation values of the first two moments of $J_z$ from the SME gives the following scalar stochastic differential equations:
\begin{eqnarray}
d\langle J_z \rangle(t)&=&\gamma \langle J_x \rangle(t) b(t)\,dt \nonumber\\
&&+ 2\sqrt{M\eta}\langle \Delta J_z^2\rangle(t) \,dW(t)\\
d\langle \Delta J_z^2\rangle(t)&=&-4M\eta\langle\Delta J_z^2\rangle^2(t)\,dt\nonumber\\
&&-i\gamma\langle[\Delta J_z^2 ,  J_y]\rangle(t)b(t)\,dt\nonumber\\
&&+ 2\sqrt{M\eta}\langle \Delta J_z^3 \rangle(t) dW(t)
\end{eqnarray}
Note that these equations are not closed because higher order moments couple to them.

At short times,  $t \ll 1/\eta M$,  we can make this set of equations closed with the following approximations.  If the spins are initially fully polarized along $x$ then, by using the evolution equation for the $x$ component, we can show $\langle J_x \rangle(t) \approx J \exp[-M t /2]$. Making the Gaussian approximation at short times, the third order terms $\langle \Delta J_z^3 \rangle$ and $-i\gamma\langle[\Delta J_z^2 , J_y] \rangle(t)b(t)$ can be neglected. The Holstein-Primakoff transformation makes it possible to derive this Gaussian approximation as an expansion in $1/J$ \cite{Holstein1940}. Both of the removed terms can be shown to be approximately $1/J\sqrt{J}$ smaller than the retained non-linear term.  Thus we can approximate the optimal solution with
\begin{eqnarray}
d\langle J_z \rangle_s(t)&=& \gamma J\exp[-M t /2] b(t)\,dt \nonumber\\
&&+ 2\sqrt{M\eta}\langle \Delta J_z^2\rangle_s(t)\,dW_s(t)\\
d\langle \Delta J_z^2\rangle_s(t)&=& -4M\eta\langle\Delta J_z^2\rangle_s^2(t)\,dt
\end{eqnarray}
where the $s$ subscript denotes the short time solution and $dW_s(t)\equiv 2 \sqrt{M\eta}[y(t) dt-\langle J_z\rangle_s (t) dt]$. Also $b(t)$ is assumed to be of a form that keeps the total state nearly pointing along x. The differential equation for the variance $\langle\Delta J_z^2\rangle_s(t)$ is now
deterministic. It can be solved to give
\begin{equation}
\langle \Delta J_z^2\rangle_s(t)=\frac{\langle \Delta
J_z^2\rangle(0)}{1+4 M \eta \langle \Delta J_z^2\rangle(0)t}\label{Equation::ShortTimeVarianceSolution}
\end{equation}
The deterministically shrinking value of $\langle \Delta J_z^2\rangle_s(t)$ represents the squeezing about the initially fluctuating value of $\langle J_z \rangle_s(t)$ as shown in the first two frames of \reffig{1}A-B. If feedback is added, then the value of $\langle J_z \rangle_s(t)$ can be zeroed via Larmor precession due to a control field along y and the same centered spin squeezed state can be prepared on every trial \cite{Thomsen2002a,Thomsen2002b,Geremia2004a}.

The resulting spin squeezed states can be used in subsequent precision measurements \cite{Wineland1994, Polzik2003}. It is also worth pointing out that a precision measurement can be performed \emph{during} the production of the conditional spin squeezing. For example, we have shown that the by properly estimating both the spin state and an unknown classical field simultaneously with continuous measurement and Kalman filtering techniques, the field estimation can be improved over conventional limits by the presence of the simultaneous squeezing \cite{Geremia2003a,Stockton2003b}.

\subsection{\label{Section::LongTime}Long time limit}

The approximations made in the previous section are no longer valid at times $t\gg1/\eta M$.  The third-order terms become non-negligible at long times, hence the variance becomes stochastic. Subsequently, other high order moments couple into the problem and we are forced to consider the stochastic differential equation for each. Eventually, any finite numbered moment description is no longer useful and it initially appears that we must resort back to the full symmetric density matrix and the SME, \refeqn{SME}, as our primary description.

Fortunately, we can take another approach and describe the state in terms of other sufficient statistics. Without a field, the only statistic of the photocurrent needed to describe the state at time $t$ is its integral, $\int_0^t y(s)\,ds$  (see \refapx{App1} or \cite{Hughston2002}). Knowing that the state is only a function of this variable and the initial state (prior information) makes the experimental design of a real time estimator experimentally convenient.  For example, we could use an analog integrator to create this sufficient statistic from the raw photocurrent, then feed it into a possibly non-linear device (like an FPGA \cite{Stockton2002}) to perform the estimation.

With the integrated photocurrent and the initial state
\begin{equation}
    |\psi (0)\rangle=\sum_{m=-J}^J c_m |m\rangle
\end{equation}
we can calculate (see \refapx{App1}) the conditional expectation value of any power of $J_z$ with the expression
\begin{eqnarray}
{\rm Tr}[J_z^k\tilde\rho(t)]&=&\sum_{m=-J}^J m^k |c_m|^2 \exp[-2M\eta m^2 t \nonumber\\
&& + 4 m M \eta \int_0^t y(s)\,ds]\label{Equation::unMom}
\end{eqnarray}
where $\tilde\rho(t)$ is the unnormalized density matrix, and setting $k=0$ represents its trace, so
\begin{equation}
\langle J_z^k \rangle (t) ={\rm Tr}[J_z^k\tilde\rho(t)]/{\rm Tr}[J_z^0\tilde\rho(t)]
\end{equation}
Consider the case when the system starts in the $x$-polarized spin-coherent state. To very good approximation (with reasonably large $J$) we can write for this state in the $z$-basis
\begin{equation}
    |c_m|^2\propto\exp\left[-\frac{m^2}{J}\right]
\end{equation}
Using these coefficients, we now have the rule for mapping the photocurrent to the expectation of $J_z$
\begin{equation}
\langle J_z \rangle (t) ={\rm Tr}[J_z^1 \tilde\rho(t)]/{\rm Tr}[J_z^0 \tilde\rho(t)]
\end{equation}
Other than the minor approximation of the initial coefficients, using this estimate is essentially the same as using solution to the full SME, so we do not give it a new subscript.

To simplify further, we can change the sums to integrals, giving
\begin{equation}
    {\rm Tr}[m^k \tilde\rho(t)] \simeq \int_{-J}^J m^k e^{-Am^2+2Bm}dm
\end{equation}
with
\begin{equation}
    A = \frac{1}{J}+2M\eta t~~~~~~~
    B = 2 M\eta\int_0^t y(s)\,ds
\end{equation}
This approximation produces an estimate
\begin{equation}
\langle J_z \rangle_i(t)= \frac{\int_{-J}^J m e^{-Am^2+2Bm}dm }{ \int_{-J}^J e^{-Am^2+2Bm}dm}
\end{equation}
that performs sub-optimally when the distribution of states becomes very narrow at long times. Interestingly, the integral approximation here numerically appears to give the same estimate as the one derived previously for short times when no field is present, i.e.
\begin{equation}
\langle J_z \rangle_i(t)=\langle J_z \rangle_s(t)
\end{equation}
This is not entirely surprising as both of these estimators ignore the discreteness of the Dicke levels. Also, at long times, it turns out that both of these estimates appear to be numerically equivalent to the simplest of all estimates: averaging the photocurrent.  In other words, one simple and intuitive approximation to the optimal $\langle J_z \rangle(t)$ would be
\begin{equation}
\langle J_z \rangle_a(t)=\frac{\int_0^t y(s)ds}{t} \label{Equation::CurrentAve}
\end{equation}
which is an estimate one might guess from the form of the photocurrent, \refeqn{Photocurrent}.  From simulation, it appears that this estimate is the same as both $\langle J_z \rangle_i(t)$ and $\langle J_z \rangle_s(t)$ for $t\gg1/\eta M$. Despite the non-optimality of these simple estimators, they perform well enough to resolve the discretization of the Dicke levels at long times. 

Unfortunately, the addition of a feedback field makes these simplified estimators inadequate at long times, and deriving simple reduced models with a field present is difficult, thus forcing us to use the full SME in our state based controller.  Despite this difficulty, during our subsequent feedback analysis we assume sufficient control bandwidth that the SME can be evolved by the observer in real time.

\begin{figure*}
\includegraphics[width=7in]{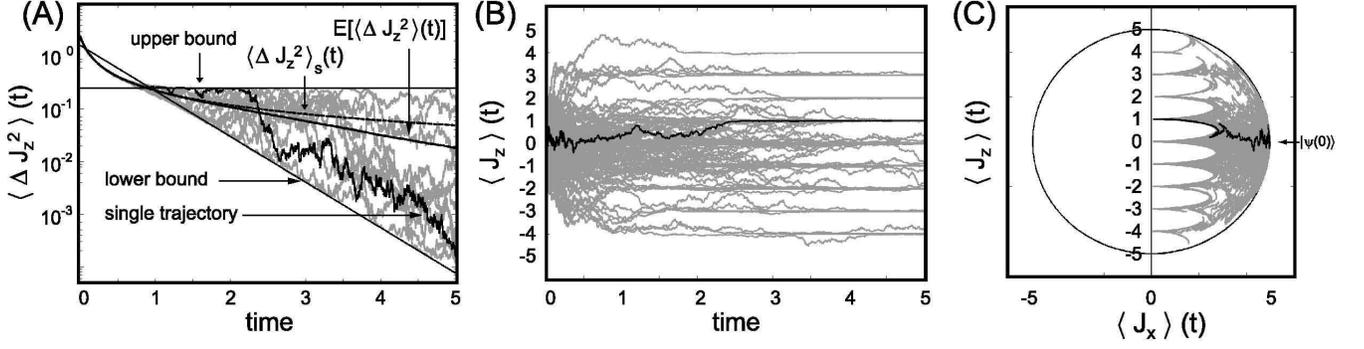} \caption{Many open-loop moment trajectories [36] of the SSE, \refeqn{SSE}. The trajectory of \reffig{1} is darkened.  (A) At short times, the evolution of the variance (shown on a log scale) is deterministic and given by $\langle \Delta J_z^2 \rangle_s(t)$.  At long times, the variances become stochastic but bounded (above by $1/4$ and below by $\exp[-2(M t-1)]/4$).  The average of all $10,000$ trajectories (only $10$ are shown) gives $\textrm{E}[\langle \Delta J_z^2\rangle(t)]$.  (B) The projective nature of the measurement is made clear by the evolution of $100$ trajectories of $\langle J_z\rangle(t)$.  The distribution of the final results is given by the first histogram of \reffig{1}A.  (C)  The evolution of the $100$ trajectories all starting in an $x$-polarized CSS.  When $\eta=1$, certain regions of Hilbert space are forbidden by the evolution.}\label{Figure::2}
\end{figure*}

\section{\label{Section::OL}MEASUREMENT EVOLUTION WITHOUT FEEDBACK}

In this section, our goal is to describe how the estimates of the last section probabilistically evolve at long times into Dicke states via observation alone. First, we discuss steady state and statistical properties of the SME, \refeqn{SME}. Then, we examine the unconditional dynamical solution with $\eta=0$ which gives the average state preparation behavior when $\eta\neq 0$. We then consider in detail how individual trajectories behave when $\eta\neq 0$. Finally, we discuss the performance of the non-optimal estimators relative to the optimal projective estimator.

\subsection{Steady states of the SME and martingale properties}

The fact that the SME eventually prepares eigenstates of $J_z$ is rather intuitive from a projection postulate perspective because $J_z$ is the quantity being measured. If we insert the pure Dicke state, $\rho = |m\rangle\langle m|$, into the SME with no Hamiltonian (or only a field along z), we find that it is a steady state, $d\rho=0$, no matter what happens with the subsequent measurement record. Of course, this does not yet prove that the state will eventually be obtained, as we have not discussed the stability of attractors in stochastic systems.

Without a field present, the SME has several convenient properties. First of all, from the evolution equation for the variance notice that the variance is a stochastic process that decreases on average.  In fact it is a super-martingale, in that for times $s \leq t$ we have
\begin{equation}
\textrm{E}_s[\langle \Delta J_z^2\rangle(t)]\leq \langle \Delta J_z^2\rangle(s)
\end{equation}
where the notation $\textrm{E}[x(t)]$ denotes the average of the stochastic variable $x(t)$ at time $t$ and the $s$ subscript represents conditional expectation given a particular stochastic trajectory up to the time $s$. Additionally, it can be shown \cite{Hughston2001} that the average variance obeys the equation
\begin{equation}
\textrm{E}[\langle \Delta J_z^2\rangle(t)]=\frac{\langle \Delta J_z^2\rangle(0)}{1+4 M \eta \langle \Delta
J_z^2\rangle(0)(t+\xi(t))}\label{Equation::AveVariance}
\end{equation}
where
\begin{eqnarray}
\xi(t)&=&\int_0^t \frac{ \textrm{E}[(\langle \Delta J_z^2\rangle(s)-\textrm{E}[\langle \Delta
J_z^2\rangle(s)])^2]}{\textrm{E}[\langle \Delta J_z^2\rangle(s)]^2}\,ds\nonumber\\
&\geq&0\label{Equation::XiEquation}
\end{eqnarray}
A more explicit solution of $\xi(t)$ is not necessarily needed as its positivity ensures that $\langle \Delta J_z^2 \rangle(t)$ stochastically approaches zero.  This implies that a Dicke state is eventually prepared.  The numerical simulation of \reffigs{1}{2} demonstrates this behavior for an initially x polarized state. As expected, $\textrm{E}[\langle \Delta J_z^2\rangle(t)]$ in \reffig{2}A appears to be less than the short time solution $\langle \Delta J_z^2\rangle_s(t)$, \refeqn{ShortTimeVarianceSolution}, at long times.

Other useful properties of the stochastic evolution are evident from the moment equations.  For example, we can show that
\begin{equation}
d\langle J_z^n\rangle = 2\sqrt{M\eta}(\langle J_z^{n+1}\rangle-\langle J_z^n\rangle\langle J_z\rangle)dW(t)
\end{equation}
for integer $n$, hence
\begin{equation}
d\textrm{E}[\langle J_z^n\rangle] = 0
\end{equation}
and for times $s \leq t$ we have the \emph{martingale} condition
\begin{equation}
\textrm{E}_s[\langle J_z^n\rangle(t)] = \langle J_z^n\rangle(s)
\end{equation}
This equation for $n=1$ gives us the useful identity
\begin{equation}
\textrm{E}[\langle J_z\rangle(t)\langle J_z\rangle(s)] =\textrm{E}[\langle J_z\rangle(s)^2]
\end{equation}
for $s \leq t$. Also, we can re-write the expression for $n=2$ as
\begin{equation}
\textrm{E}_s[\langle J_z\rangle(t)^2+\langle \Delta
J_z^2\rangle(t)] = \langle J_z\rangle(s)^2+\langle \Delta J_z^2\rangle(s)
\end{equation}
This implies a sort of conservation of uncertainty as the diffusion in the mean, shown in \reffig{1}B, makes up for the decreasing value of the variance.

\subsection{$\eta=0$}

It is insightful to examine the behavior of the master equation with $\eta=0$ which corresponds to ignoring the measurement results and turns the SME \refeqn{SME} into a deterministic unconditional master equation.  We continue to consider only those initial states that are polarized. This is because these states are experimentally accessible (via optical pumping) and provide some degree of selectivity for the final prepared state. To see this, let us consider a spin-$1/2$ ensemble polarized in the $x-z$ plane, making angle $\theta$ with the positive $z$-axis, such that
\begin{eqnarray}
\langle  J_x \rangle(0)&=&\sin(\theta)N/2\\
\langle  J_y \rangle(0)&=&0\nonumber\\
\langle  J_z \rangle(0)&=&\cos(\theta)N/2\nonumber\\
\langle \Delta J_x^2 \rangle (0)  &=& \cos^2(\theta)N/4\nonumber\\
\langle \Delta J_y^2 \rangle (0)  &=& N/4\nonumber\\
\langle \Delta J_z^2 \rangle (0)  &=& \sin^2(\theta)N/4\nonumber
\end{eqnarray}
Solving the unconditional moment equations, and labelling them with $u$ subscripts, we get
\begin{eqnarray}
\langle  J_x \rangle_u(t)&=&\sin(\theta)\exp(-Mt/2)N/2\\
\langle  J_y \rangle_u(t)&=&0\nonumber\\
\langle  J_z \rangle_u(t)&=&\cos(\theta)N/2\nonumber\\
\langle \Delta J_x^2 \rangle_u (t)  &=& \sin^2(\theta)[N^2-N-2N^2\exp(-Mt)\nonumber\\
&&+(N^2-N)\exp(-2Mt)]/8+N/4\nonumber\\
&\rightarrow&\sin^2(\theta)(N^2-N)/8+N/4\nonumber\\
\langle \Delta J_y^2  \rangle_u (t) &=&  \sin^2(\theta)[N^2-N\nonumber\\
&&+(N-N^2)\exp(-2Mt)]/8+N/4\nonumber\\
&\rightarrow&\sin^2(\theta)(N^2-N)/8+N/4\nonumber\\
\langle \Delta J_z^2  \rangle_u (t) &=&  \sin^2(\theta) N/4\nonumber
\end{eqnarray}
Note that, because the unconditional solutions represent the average of the conditional solution, i.e.
$\rho_u(t)=\textrm{E}[\rho(t)]$, we have
\begin{equation}
\textrm{E}[\langle  J_z \rangle(t)]=\langle  J_z
\rangle_u(t)=\langle  J_z \rangle(0)=\cos(\theta)N/2
\end{equation}
This also follows from the martingale condition for $\langle  J_z
\rangle(t)$.  From the martingale condition for $\langle  J_z^2
\rangle(t)$ we get
\begin{eqnarray}
\textrm{E}[(\langle J_z \rangle(t)&-&\textrm{E}[\langle J_z \rangle(t)])^2]\nonumber\\
&=&\langle \Delta J_z^2 \rangle(0)-\textrm{E}[\langle \Delta J_z^2 \rangle(t)] \\
&\rightarrow& \langle \Delta J_z^2 \rangle(0)=\sin^2(\theta) N/4\nonumber
\end{eqnarray}
Thus, when $0 < \eta \leq 1$, we expect the final random conditional Dicke state on a given trial to fall within the initial $z$ distribution. Given $\theta$, the distribution will have spread $|\sin(\theta)|\sqrt{N}/2$ about the value $\cos(\theta)N/2$. Although the final state is generally random, starting with a polarized state clearly gives us some degree of selectivity for the final Dicke state because $\sqrt{N}\ll N$.

\subsection{$0 < \eta \leq 1$}

When $\eta\neq 0$, the measurement record is used to condition the state, and we can determine which Dicke state the system diffuses into.  Given the task of preparing the state $|m_d\rangle$, the above analysis suggests the following experimental procedure. First, polarize the ensemble (via optical pumping) into an unentangled coherent state along any direction. Then rotate the spin vector (with a magnetic field) so that the $z$ component is approximately equal to $m_d$.  Finally, continuously measure $z$ until a time $t\gg1/\eta M$. The final estimate will be a random Dicke state in the neighborhood of $m_d$.  When the trial is repeated, the final states will make up a distribution described by the initial moments of $J_z$ ($\langle  J_z \rangle(0)$, $\langle \Delta J_z^2 \rangle(0)$, ...).  To reduce the effects of stray field fluctuations and gradients, a strong holding field could be applied along the $z$-axis.  Because this Hamiltonian commutes with the observable $J_z$, the final open-loop measurement results would be unchanged. 

This process (with zero field) is shown schematically in \reffig{1} for $m_d=0$ where the initial state is polarized along $x$. Because $\langle  J_z \rangle(0)=0$, the final state with the highest probability is the entangled Dicke state $m_d=0$.  In contrast, if $\langle J_z \rangle(0)=J$ the state would start in an unentangled CSS polarized along $z$ and would not subsequently evolve.

One way of characterizing how close the state is to a Dicke state is through the variance, $\langle \Delta J_z^2 \rangle (t)$. \reffig{2}A displays many trajectories for the variance as a function of time.  For times $t\ll1/\eta M$ the variance is approximately deterministic and obeys the short time solution of \refeqn{ShortTimeVarianceSolution}. During this period, the mean $\langle J_x \rangle(t)$ is decreasing at rate $M/2$.  Before this mean has completely disappeared, a conditional spin squeezed state is created. However, for larger times the mean and the variance stochastically approach zero, and the state, while still entangled, no longer satisfies the spin-squeezing criterion \cite{Stockton2003a}.

There are several features to notice about the approach to a Dicke state that are evident in \reffigs{1}{2}. The variance at time $t=1/\eta M$ is already of order unity. Thus, at this point, only a few neighboring $m$ levels contain any population, as can be seen in \reffig{1}C. Also, it can be numerically shown that, for $x$-polarized initial states, the diffusion of the variance at long times $t\gg1/\eta M$ is bounded above and below by
\begin{equation}
\exp[-2(\eta M t-1)]/4 < \langle \Delta J_z^2 \rangle (t) \leq 1/4\label{Equation::LongTimeBounds}
\end{equation}
which is evident from \reffig{2}A.  These facts indicate that the population is divided among at most two levels at long times which `compete' to be the final winner.  If we assume that only two neighboring levels are occupied and apply the SSE (with $\eta=1$), the probability, $p$, to be in one level obeys the stochastic equation
\begin{equation}
dp = -2 M p (1-p)\,dW(t)
\end{equation}
and the variance takes the form $\langle \Delta J_z^2 \rangle (t)=p (1-p)$.  As simple as it looks, this stochastic differential equation (SDE) is not analytically solvable \cite{Oksendal1998, Gardiner2002}. The maximum variance is $1/4$ and it can be shown that, for $p\equiv1-\epsilon$, with $\epsilon$ small, the lower bound is of the exponential form stated above, so the two-level assumption seems to be a good one. The fact that occupied Hilbert space becomes small at long times is also evident in \reffig{2}C, where the allowed states are seen to be excluded from certain regions when $\eta=1$.  The arc-like boundaries of the forbidden space are where the two level competition occurs.

In practice, an experimentalist does not always have an infinite amount of time to prepare a state.  Eventually spontaneous emission and other decoherence effects will destroy the dispersive QND approximation that the present analysis is based upon.  Suppose our task was to prepare a Dicke state with, on average, a desired uncertainty, $\langle \Delta J_z^2\rangle_d\ll 1$, such that one level was distinguishable from the next.  From \refeqn{AveVariance}, we see that the time that it would take to do this on average is given by
\begin{equation}
t_d = \left[\frac{1}{\langle \Delta J_z^2 \rangle_d }-\frac{1}{\langle \Delta J_z^2 \rangle(0)}\right]/4M\eta
\label{Equation::EndTime}
\end{equation}
Thus if $\langle \Delta J_z^2\rangle_d\ll 1$ is our goal, then $t_d$ is how long the state must remain coherent.  The larger $\langle \Delta J_z^2 \rangle(0)$ is the more entangled the final states are likely to be ($m\approx0$) \cite{Stockton2003a}, hence, by \refeqn{EndTime}, the longer it takes to prepare the state for a given $\langle \Delta J_z^2\rangle_d$.  Hence, we arrive at the intuitively satisfying conclusion that conditional measurement produces entangled states more slowly than unentangled states. Of course, \refeqn{EndTime} is an average performance limit. In a best case scenario, the variance would attain the lower bound of \refeqn{LongTimeBounds} where the state reduction happens exponentially fast.

\subsection{Performance of sub-optimal estimators}

Now we consider the performance of the sub-optimal estimators discussed previously, in particular the current average $\langle J_z \rangle_a(t)$ of \refeqn{CurrentAve}. It makes sense to associate the overall `error' of this estimator, denoted $V_a$, to be the average squared distance of the estimator from the optimal estimator \emph{plus} the average uncertainty of the optimal estimator itself, $\textrm{E}[\langle \Delta J_z^2\rangle(t)]$. Using the martingale properties of the optimal estimate and the definition of the photocurrent gives this quantity as 
\begin{eqnarray}
V_a&\equiv&\textrm{E}[(\langle J_z \rangle_a(t)-\langle J_z \rangle (t))^2] +\textrm{E}[\langle \Delta J_z^2 \rangle(t) ]\nonumber\\
&=& \frac{1}{4 M\eta t}
\end{eqnarray}
This is just the error in estimating a constant masked by additive white noise with the same signal to noise ratio \cite{Stockton2003b}.  The optimal estimator is better than this sub-optimal estimator at long times only through the quantity $\xi(t)$, \refeqn{XiEquation}.

In the open-loop experimental procedure described at the beginning of the last section, the above observation indicates that we can replace the optimal estimator with the photocurrent average and still resolve the projective behavior (given sufficient elimination of extraneous noise).  The price paid for the simplicity of the averaging estimator is that it converges more slowly and it only works when a field is not present (hence without control).

\section{\label{Section::CL}CLOSED LOOP EVOLUTION}

The primary problem with the open-loop state preparation scheme (and other approaches \cite{Duan2003, Molmer2003, Unanyan2002}) is that it is probabilistic. For a single measurement, there exists some degree of control, by adjusting the initial angle of rotation $\theta$, but the final state is a priori unpredictable within the variance of the initial state. In this section, we show that the state preparation can be made deterministic with the use of feedback. Just as the control scheme of \cite{Thomsen2002a, Thomsen2002b} produces deterministically centered spin squeezed states, we present a simple feedback controller that will prepare the same desired Dicke state (particularly $m_d=0$) on every measurement trial.

We choose to work with $y$-axis magnetic field actuator corresponding to the Hamiltonian, $H(t)=\gamma b(t) J_y$.  If the CSS initial state begins in the $x-z$ plane this will ensure that the vector $\langle \vec{J} \rangle(t)$ remains in this plane. This actuator is natural for the control of spin-squeezed states at short times, where the linear moments of $\langle \vec{J} \rangle(t)$ are large and allow intuitive rotation of the spin vector.  However, at long times the field will mostly be affecting non-linear terms in the moment expansion and the dynamics are less intuitive as can be seen by the structure near the $z$-axis in \reffig{2}C.  Still, we continue to give ourselves only these rotations to work with as they are the most experimentally accessible actuation variable.

In principle, the fact that Dicke states can be prepared deterministically with feedback should not be surprising.  Given the aforementioned characteristics of the non-controlled measurement one could imagine preparing a particular state by \emph{alternating} measurement and control periods.  For example, an initial measurement (lasting for a time $\Delta t \ll 1/\eta M$) would determine the fluctuation of $\langle J_z \rangle$ while the uncertainty $\langle \Delta J_z^2 \rangle$ simultaneously decreased (on average).  Then the measurement would be turned off and the state would be rotated with a control field to `zero' the conditional quantity $\langle J_z \rangle-m_d$ (if preparing $|m_d\rangle$). The process of alternating measurement and control could then be repeated and would eventually clamp down on the desired state. Notice that, unlike the preparation of spin squeezed states \cite{Thomsen2002a, Thomsen2002b}, this procedure could not be performed with a \emph{single} measurement and control cycle. In other words, if we measure for a time $t\gg 1/\eta M$, and prepare a probabilistic Dicke state, then a single post-measurement rotation cannot prepare a different desired Dicke state in the same basis. 

With this intuitive picture in mind, now consider the continuous limit of this process, where the measurement and control are performed simultaneously.  We wish to find a mapping from the photocurrent history to the control field that prepares our state of interest in a satisfactory manner on \emph{every} trial. For simplicity, we work with $\eta=1$ and use the SSE of \refeqn{SSE} for all simulations [36]. In selecting a controller, we could choose one of several strategies, including either direct current feedback or a feedback rule based on the state (i.e., what has been called Markovian and Bayesian feedback respectively \cite{Wiseman2002, Doherty1999}).  While direct current feedback possesses certain advantages, mainly simplicity that allows practical implementation, and is capable of working adequately at short times, any constant gain procedure would never prepare a Dicke state with confidence.  If the current is directly fed back, a finite amount of noise will unnecessarily drive the system away from its target, even if the state starts there.  Of course the gain could be ramped to zero in time, but unlike the short time case, it is not clear how to tailor the gain intelligently.  

Another alternative would be to prepare a spin squeezed state with this approach and then turn off the feedback at some intermediate time.  This would certainly enhance the probability of obtaining a certain Dicke state, but the process would remain probabilistic to some degree.  For these reasons, we continue considering only state based feedback, despite the fact that updating the state in real time is experimentally challenging.

\begin{figure*}
\includegraphics[width=7in]{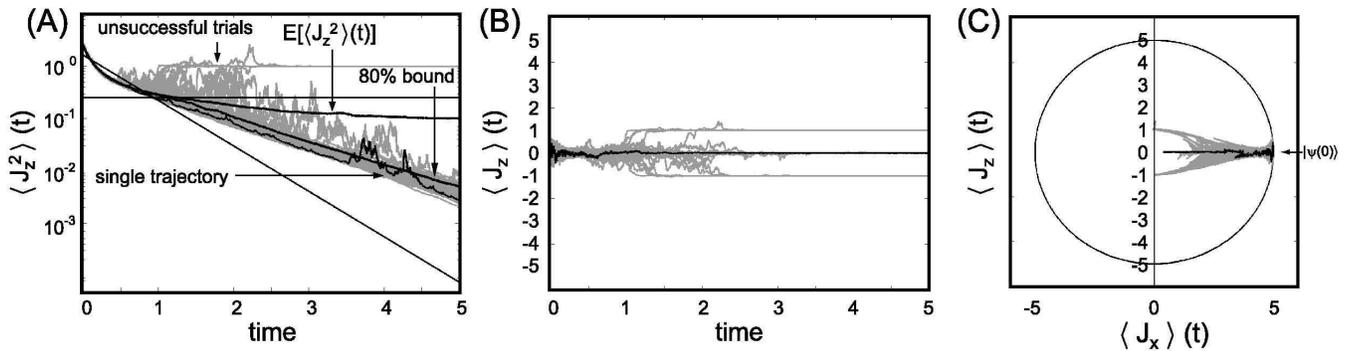} \caption{One hundred closed-loop moment trajectories [36] of the SSE with feedback law $b(t)=\lambda \langle J_x J_z + J_z J_x \rangle(t)/2$ and $\lambda=10$ chosen from numerical considerations. (A-B) If the control is successful the quantity $\langle J_z^2\rangle(t)$ should go to zero on every trial.  For this controller the number of successful trajectories is increased significantly (from $25$ to $90$ percent), but the remaining fraction is attracted to neighboring fixed points, causing the mean $\textrm E[\langle J_z^2 \rangle(t)]$ to saturate at a non-zero value.  Although the successful fraction converges exponentially, the fastest converging trajectories are slower than in the open loop case.  This is evident in (C) as the converging trajectories have visibly not yet reached $\langle J_x \rangle=0$ at time $t=5$. }\label{Figure::3}
\end{figure*}

\subsection{Defining a cost}

A useful first step in the design of any controller is to define the quantity that the ideal controller should minimize: the cost function. For example, consider a state preparation application where the controller aims to produce the desired target state $|\psi_d\rangle$.  In this case, one possible cost function is the quantity
\begin{equation}
U_f\equiv1-\langle \psi_d | \rho | \psi_d \rangle \geq 0
\end{equation}
evaluated at the stopping time, which is zero iff the fidelity of the state with respect to the target is unity. In the current application, where we desire a final Dicke state $|m_d\rangle$ we wish to minimize a different quantity
\begin{eqnarray}
U&\equiv&(\langle J_z \rangle-m_d)^2 + \langle \Delta J_z^2 \rangle\nonumber\\
&=&\Sigma_m \langle m | \rho | m \rangle^2 (m-m_d)^2\nonumber\\
&\geq& 0
\end{eqnarray}
which is zero iff $\rho=|m_d\rangle\langle m_d|$. Notice that $U$ gives a higher penalty than $U_f$ to states that are largely supported by Dicke states far removed from the target.  In general, $U$ will evolve stochastically and we may be more interested in the mean behavior, denoted $\textrm E[U]$.  In the uncontrolled case, it can be shown that this quantity remains constant, $\textrm E[U(t)] = U(0)$.  For the controlled case, we wish for $\textrm E[U] \rightarrow 0$ as time advances, which, because $U\geq 0$, implies that every trajectory approaches the target state $|m_d\rangle$.

In general, the cost function could also include an integral of the quantity $U(t)$ instead of just the final value.  As in classical control theory \cite{Stockton2003b}, it is also practical to include a function of $b(t)$ in the cost as a way of expressing our experimental feedback gain and bandwidth constraints.  Analytically proceeding in this way by optimizing the average cost is too difficult for the current problem, but, with this perspective in mind, we proceed by proposing controllers according to related considerations.

\subsection{Control law 1}

Now consider the average evolution of the above cost function, which is given by
\begin{eqnarray}
d\textrm E[U(t)]=&\nonumber\\
-2\gamma \textrm E&\left[b(t)\left(\frac{\langle J_x J_z + J_z J_x \rangle(t)}{2}-m_d \langle J_x\rangle(t)\right)\right]dt
\end{eqnarray}
Because we want this function to continuously decrease, the right hand side should be negative at all times.  If we have full access to the density matrix, and minimal feedback delay, we could use the controller
\begin{equation}
b_1(t)=\lambda \left(\frac{\langle J_x J_z + J_z J_x \rangle(t)}{2}-m_d \langle J_x\rangle(t)\right)
\end{equation}
where $\lambda$ is a constant positive gain factor.  This law guarantees that $d\textrm E[U(t)] \leq 0$. Still, this does not yet prove that $U=0$ is obtained because $d\textrm E[U(t)] = 0$ for states other than the target state. Furthermore, even with this control law applied, all Dicke states \emph{remain} fixed points.  

Regardless of these issues, we proceed by analyzing the performance of this control law numerically with $m_d=0$.  In principle, the gain could be chosen arbitrarily large.  Here we choose to work with a gain that is large enough to be effective but small enough to keep the numerical simulation results valid [36].  The choice of a limited gain is a necessity in both simulation and experiment, thus we wish to find a control law that works within this constraint.  For the parameters used in our simulation, we use a gain of $\lambda=10$ which produces the results shown in \reffig{3}.   

\begin{figure*}
\includegraphics[width=7in]{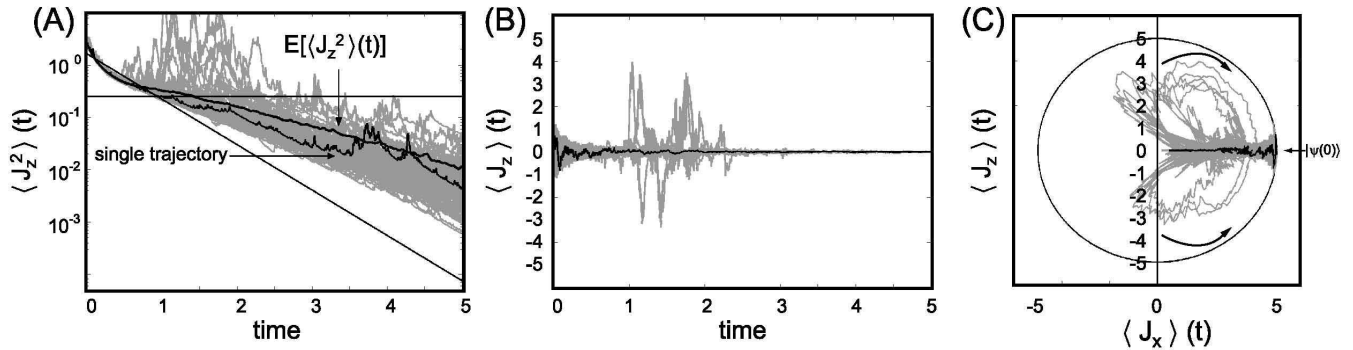} \caption{One hundred closed-loop moment trajectories [36] of the SSE with feedback law $b(t)=\lambda \langle J_z \rangle(t)$ and $\lambda=10$ chosen from numerical considerations.  (A) The average over 10,000 trajectories suggests that with this control law the mean $\textrm E[\langle J_z^2 \rangle(t)]$ descends to zero exponentially and the target state is deterministically prepared.  (B) Despite a number of early excursions, all 100 trajectories shown converge to the desired value of $m=0$.  (C)  Those trajectories that do not descend to the goal directly (about 10 of 100) are recycled and rotated back into the attractive region of the target state.  Again, the control slightly compromises the best-case convergence rate and the trajectories have a non-zero (but still decreasing) $\langle J_x \rangle$ at $t=5$.}\label{Figure::4}
\end{figure*}

In \reffig{3}A, we now plot the figure of merit for $m_d=0$, $U(t)=\langle J_z^2 \rangle(t)$. In open loop configuration, only $25\%$ of all trajectories are attracted to $m=0$, whereas with this controller the percentage reaches $90\%$.  Furthermore, most of these trajectories approach the state at an exponential rate close to $M$, as indicated by the curve under which $80\%$ of the trajectories lie.  Interestingly, this is at the expense of those trajectories that in open-loop approached the target state at an exponential rate of $2M$.  There is a trade-off by which the control slightly compromises the convergence of the best case trajectories.

Unfortunately, because all other Dicke states are still fixed points of the controlled SSE and the gain is finite, a small fraction ($10\%$) of trajectories are attracted to those states neighboring the target state.  Thus this controller does not appear to deterministically prepare all trajectories into the target state and the mean $\textrm E[\langle J_z^2 \rangle(t)]$ flattens at a level determined by the unsuccessful fraction of trials.

\subsection{Control law 2}

The obvious solution to the above problem is to try a controller that ensures the target state is the \emph{only} fixed point of the SME/SSE.  In this section we propose the control law
\begin{equation}
b_2(t)=\lambda ( \langle J_z\rangle(t) - m_d )
\end{equation}
for which the state $|m_d\rangle$ is the only fixed point.  However, unlike $b_1(t)$ this controller lacks the x symmetry that ensures $d\textrm E[U(t)] \leq 0$.  Also, while the symmetry of $b_1(t)$ will allow it to lock to both sides of the Bloch sphere, $b_2(t)$ will only lock to one side of the sphere.  

Again, we proceed by numerically analyzing the performance of this controller for $m_d=0$, with the results displayed in \reffig{4}.  The gain is chosen in the same manner as before, which leads to the same reasonable choice of $\lambda=10$.  In \reffig{4}C the fundamental nature of the dynamics can be seen.  Close to $90\%$ of the trajectories are directly transported towards the target state, but the remaining fraction `misses' on the first pass.  Instead of being attracted towards other fixed points though, this unsuccessful fraction is \emph{recycled} and rotated back onto the positive $x$-axis where they can re-attempt convergence onto the target state.  These large excursions can be seen in \reffig{4}A-B as well, but they do not appear to dominate the net flow. The average of $10,000$ trajectories gives a quantity $\textrm E[\langle J_z^2 \rangle (t)]$ which appears to exponentially descend towards zero, implying that the state preparation has been made deterministic.  As with the control of $b_1(t)$ there is again a trade-off: the trajectories that previously descended at the exponential rate of $2M$ converge more slowly, but still exponentially.

\section{\label{Section::Conclusion}Conclusion}

The purpose of this paper is to demonstrate the fact that the process of continuous projective measurement can be made deterministic with a theoretically simple and intuitive state-based control law.  In the context of an atomic spin ensemble, the resulting Dicke states are highly entangled and otherwise difficult to reliably produce from an initially unentangled state.  

However, there is much work to be done in the general field of quantum state estimation and control, of which this is one example. In this pursuit, it is helpful to utilize and adapt methods from the developed fields of classical stochastic estimation and control theory.  In \cite{vanHandel2004}, for example, the problem of this paper is considered for a single spin with greater emphasis on technical notions of stochastic stability and convergence.  Ultimately, we would like to discover constructive methods for deriving optimal control laws given a cost function and realistic actuators.

Even with an optimal control law in hand, there is no guarantee that experimental implementation will be possible.  Any analysis should incorporate, among other constraints, non-unity detection efficiencies and finite controller resources (bandwidth, memory, etc).  For experimental application of quantum feedback, the controller complexity needs to be reduced to the point where the delay is minimal compared to other dynamical time-scales \cite{Stockton2002}. As in classical control, effective model reduction techniques are indispensable when it comes to implementation.

Despite these difficulties, the increasing number of physical systems that can be measured reliably at the quantum limit will surely hasten the effort to solve many of these technical challenges.  By respecting the physical basis of measurement dynamics, experimentalists will be able to more efficiently use measurement itself, in tandem with more traditional techniques, to actuate quantum systems into desirable states.

\acknowledgments

This work was supported by the NSF (PHY-9987541, EIA-0086038), the ONR (N00014-00-1-0479), the ARO (DAAD19-03-1-0073), and the Caltech MURI Center for Quantum Networks (DAAD19-00-1-0374). JKS acknowledges a Hertz fellowship. We thank Andrew Doherty, JM Geremia, and Paige Randall for useful discussions.  Additional information is available at http://minty.caltech.edu/Ensemble.

\appendix

\section{\label{Appendix::App1}Solution of the SME without a field}

An explicit solution to the SME, \refeqn{SME}, can easily be found in the case $H(t)=0$.   First, the SME is rewritten as
\begin{equation}
\label{Equation::linSME}
	d\tilde\rho(t)=\mathcal{D}[\sqrt{M}J_z]\tilde\rho(t)dt+2M\eta\,(J_z\tilde\rho(t)+\tilde\rho(t)J_z)\,y(t)dt
\end{equation}
This equation, known as the {\it unnormalized} or {\it linear} SME, is equivalent to \refeqn{SME} with the identification
\begin{equation}
	\rho(t)=\tilde\rho(t)/{\rm Tr}[\tilde\rho(t)]
\end{equation}
Introducing the notation
\begin{equation}
\begin{split}
	\mathcal{G}_1\tilde\rho&=J_z\tilde\rho J_z \\
	\mathcal{G}_2\tilde\rho&=J_z^2\tilde\rho+\tilde\rho J_z^2\\
	\mathcal{G}_3\tilde\rho&=J_z\tilde\rho+\tilde\rho J_z
\end{split}
\end{equation}
\refeqn{linSME} can be written in the more suggestive form
\begin{equation}
d\tilde\rho(t) = M(\mathcal{G}_1-\tfrac{1}{2}\mathcal{G}_2)\tilde\rho(t)dt+2M\eta\,\mathcal{G}_3\tilde\rho(t)\,y(t)dt\label{Equation::linSMEsugg}
\end{equation}
Now note that \refeqn{linSMEsugg} is a linear It\^o stochastic differential equation (SDE) \cite{Oksendal1998} for $\tilde\rho(t)$, and moreover $\mathcal{G}_{1,2,3}$ all commute with each other in the 
sense that $\mathcal{G}_i\mathcal{G}_j\tilde\rho=\mathcal{G}_j\mathcal{G}_i\tilde\rho$.
Such SDEs have a simple explicit solution \cite{Gardiner2002}
\begin{eqnarray}
\label{Equation::linSMEsol}
\tilde\rho(t)&=&\exp[ (M(1-\eta)\mathcal{G}_1 -M(1+\eta)\mathcal{G}_2/2)t \nonumber\\
&&+2M\eta\,\mathcal{G}_3\int_0^t y(s)ds ]\tilde\rho(0)
\end{eqnarray}
as is easily verified by taking the time derivative of this expression, 
where care must be taken to use It\^o's rule for the stochastic term.

Now consider an initial pure state of the form
\begin{equation}
	|\psi(0)\rangle = \sum_{m=-J}^J c_m|m\rangle
\end{equation}
The associated initial density matrix is then
\begin{equation}
	\tilde\rho(0)=|\psi(0)\rangle\langle\psi(0)|=
		\sum_{m,m'=-J}^J c_mc_{m'}^*|m\rangle\langle m'|
\end{equation}
Substituting into \refeqn{linSMEsol} gives
\begin{equation}
\begin{split}
	\tilde\rho(t)=
	\sum_{m,m'=-J}^J &c_mc_{m'}^*
	\exp[		(M(1-\eta)mm' \\ 
		& -\tfrac{1}{2}M(1+\eta)(m^2+(m')^2))t \\
		& +2M\eta\,(m+m')\int_0^t y(s)ds] \,|m\rangle\langle m'|
\end{split}
\end{equation}
Hence 
\begin{eqnarray}
{\rm Tr}[J_z^k\tilde\rho(t)]&=&\sum_{m=-J}^J m^k |c_m|^2 \exp[-2M\eta m^2 t \nonumber\\
&& + 4 m M \eta \int_0^t y(s)\,ds]
\end{eqnarray}
which is the result used in the text, \refeqn{unMom}.

\bibliography{Paper}

\begin{thebibliography}{35}
\expandafter\ifx\csname natexlab\endcsname\relax\def\natexlab#1{#1}\fi
\expandafter\ifx\csname bibnamefont\endcsname\relax
  \def\bibnamefont#1{#1}\fi
\expandafter\ifx\csname bibfnamefont\endcsname\relax
  \def\bibfnamefont#1{#1}\fi
\expandafter\ifx\csname citenamefont\endcsname\relax
  \def\citenamefont#1{#1}\fi
\expandafter\ifx\csname url\endcsname\relax
  \def\url#1{\texttt{#1}}\fi
\expandafter\ifx\csname urlprefix\endcsname\relax\def\urlprefix{URL }\fi
\providecommand{\bibinfo}[2]{#2}
\providecommand{\eprint}[2][]{\url{#2}}

\bibitem[{\citenamefont{Wiseman}(1996)}]{Wiseman1996}
\bibinfo{author}{\bibfnamefont{H.~M.} \bibnamefont{Wiseman}},
  \bibinfo{journal}{Quant. Semiclass. Opt.} \textbf{\bibinfo{volume}{8}},
  \bibinfo{pages}{205} (\bibinfo{year}{1996}).

\bibitem[{\citenamefont{Wiseman}(1994)}]{Wiseman1994}
\bibinfo{author}{\bibfnamefont{H.~M.} \bibnamefont{Wiseman}},
  \bibinfo{journal}{Phys. Rev. A} \textbf{\bibinfo{volume}{49}},
  \bibinfo{pages}{2133} (\bibinfo{year}{1994}).

\bibitem[{\citenamefont{Dicke}(1954)}]{Dicke1954}
\bibinfo{author}{\bibfnamefont{R.~H.} \bibnamefont{Dicke}},
  \bibinfo{journal}{Phys. Rev.} \textbf{\bibinfo{volume}{93}},
  \bibinfo{pages}{99} (\bibinfo{year}{1954}).

\bibitem[{\citenamefont{Mandel and Wolf}(1997)}]{Mandel1995}
\bibinfo{author}{\bibfnamefont{L.}~\bibnamefont{Mandel}} \bibnamefont{and}
  \bibinfo{author}{\bibfnamefont{E.}~\bibnamefont{Wolf}},
  \emph{\bibinfo{title}{Optical Coherence and Quantum Optics}}
  (\bibinfo{publisher}{Cambridge University Press},
  \bibinfo{address}{Cambridge, United Kingdom}, \bibinfo{year}{1997}).

\bibitem[{\citenamefont{Adler et~al.}(2001)\citenamefont{Adler, Brody, Brun,
  and Hughston}}]{Hughston2001}
\bibinfo{author}{\bibfnamefont{S.~L.} \bibnamefont{Adler}},
  \bibinfo{author}{\bibfnamefont{D.~C.} \bibnamefont{Brody}},
  \bibinfo{author}{\bibfnamefont{T.~A.} \bibnamefont{Brun}}, \bibnamefont{and}
  \bibinfo{author}{\bibfnamefont{L.~P.} \bibnamefont{Hughston}},
  \bibinfo{journal}{J. Phys. A} \textbf{\bibinfo{volume}{34}},
  \bibinfo{pages}{8795} (\bibinfo{year}{2001}).

\bibitem[{\citenamefont{Thomsen
  et~al.}(2002{\natexlab{a}})\citenamefont{Thomsen, Mancini, and
  Wiseman}}]{Thomsen2002a}
\bibinfo{author}{\bibfnamefont{L.~K.} \bibnamefont{Thomsen}},
  \bibinfo{author}{\bibfnamefont{S.}~\bibnamefont{Mancini}}, \bibnamefont{and}
  \bibinfo{author}{\bibfnamefont{H.~M.} \bibnamefont{Wiseman}},
  \bibinfo{journal}{Phys. Rev. A} \textbf{\bibinfo{volume}{65}},
  \bibinfo{pages}{061801} (\bibinfo{year}{2002}{\natexlab{a}}).

\bibitem[{\citenamefont{Thomsen
  et~al.}(2002{\natexlab{b}})\citenamefont{Thomsen, Mancini, and
  Wiseman}}]{Thomsen2002b}
\bibinfo{author}{\bibfnamefont{L.~K.} \bibnamefont{Thomsen}},
  \bibinfo{author}{\bibfnamefont{S.}~\bibnamefont{Mancini}}, \bibnamefont{and}
  \bibinfo{author}{\bibfnamefont{H.~M.} \bibnamefont{Wiseman}},
  \bibinfo{journal}{J. Phys. B: At. Mol. Opt. Phys.}
  \textbf{\bibinfo{volume}{35}}, \bibinfo{pages}{4937}
  (\bibinfo{year}{2002}{\natexlab{b}}).

\bibitem[{\citenamefont{Kuzmich et~al.}(2000)\citenamefont{Kuzmich, Mandel, and
  Bigelow}}]{Kuzmich2000}
\bibinfo{author}{\bibfnamefont{A.}~\bibnamefont{Kuzmich}},
  \bibinfo{author}{\bibfnamefont{L.}~\bibnamefont{Mandel}}, \bibnamefont{and}
  \bibinfo{author}{\bibfnamefont{N.~P.} \bibnamefont{Bigelow}},
  \bibinfo{journal}{Phys. Rev. Lett.} \textbf{\bibinfo{volume}{85}},
  \bibinfo{pages}{1594} (\bibinfo{year}{2000}).

\bibitem[{\citenamefont{Geremia
  et~al.}(2003{\natexlab{a}})\citenamefont{Geremia, Stockton, and
  Mabuchi}}]{Geremia2004a}
\bibinfo{author}{\bibfnamefont{J.~M.} \bibnamefont{Geremia}},
  \bibinfo{author}{\bibfnamefont{J.~K.} \bibnamefont{Stockton}},
  \bibnamefont{and} \bibinfo{author}{\bibfnamefont{H.}~\bibnamefont{Mabuchi}},
  \bibinfo{journal}{unpublished}  (\bibinfo{year}{2003}{\natexlab{a}}).

\bibitem[{\citenamefont{Wineland et~al.}(1994)\citenamefont{Wineland,
  Bollinger, Itano, and Heinzen}}]{Wineland1994}
\bibinfo{author}{\bibfnamefont{D.~J.} \bibnamefont{Wineland}},
  \bibinfo{author}{\bibfnamefont{J.~J.} \bibnamefont{Bollinger}},
  \bibinfo{author}{\bibfnamefont{W.~M.} \bibnamefont{Itano}}, \bibnamefont{and}
  \bibinfo{author}{\bibfnamefont{D.~J.} \bibnamefont{Heinzen}},
  \bibinfo{journal}{Phys. Rev. A} \textbf{\bibinfo{volume}{50}},
  \bibinfo{pages}{67–88} (\bibinfo{year}{1994}).

\bibitem[{\citenamefont{Oblak et~al.}(2003)\citenamefont{Oblak, Mikkelsen,
  Tittel, Vershovski, andPlamen G.~Petrov, Alzar, and Polzik}}]{Polzik2003}
\bibinfo{author}{\bibfnamefont{D.}~\bibnamefont{Oblak}},
  \bibinfo{author}{\bibfnamefont{J.~K.} \bibnamefont{Mikkelsen}},
  \bibinfo{author}{\bibfnamefont{W.}~\bibnamefont{Tittel}},
  \bibinfo{author}{\bibfnamefont{A.~K.} \bibnamefont{Vershovski}},
  \bibinfo{author}{\bibfnamefont{J.~L.~S.} \bibnamefont{andPlamen G.~Petrov}},
  \bibinfo{author}{\bibfnamefont{C.~L.~G.} \bibnamefont{Alzar}},
  \bibnamefont{and} \bibinfo{author}{\bibfnamefont{E.~S.}
  \bibnamefont{Polzik}}, \bibinfo{journal}{quant-ph/031216}
  (\bibinfo{year}{2003}).

\bibitem[{\citenamefont{Stockton
  et~al.}(2003{\natexlab{a}})\citenamefont{Stockton, Geremia, Doherty, and
  Mabuchi}}]{Stockton2003a}
\bibinfo{author}{\bibfnamefont{J.~K.} \bibnamefont{Stockton}},
  \bibinfo{author}{\bibfnamefont{J.}~\bibnamefont{Geremia}},
  \bibinfo{author}{\bibfnamefont{A.~C.} \bibnamefont{Doherty}},
  \bibnamefont{and} \bibinfo{author}{\bibfnamefont{H.}~\bibnamefont{Mabuchi}},
  \bibinfo{journal}{Phys. Rev. A} \textbf{\bibinfo{volume}{67}},
  \bibinfo{pages}{022112} (\bibinfo{year}{2003}{\natexlab{a}}).

\bibitem[{\citenamefont{Duan and Kimble}(2003)}]{Duan2003}
\bibinfo{author}{\bibfnamefont{L.~M.} \bibnamefont{Duan}} \bibnamefont{and}
  \bibinfo{author}{\bibfnamefont{H.~J.} \bibnamefont{Kimble}},
  \bibinfo{journal}{Phys. Rev. Lett.} \textbf{\bibinfo{volume}{90}},
  \bibinfo{pages}{253601} (\bibinfo{year}{2003}).

\bibitem[{\citenamefont{S$\o$rensen and M$\o$lmer}(2003)}]{Molmer2003}
\bibinfo{author}{\bibfnamefont{A.~S.} \bibnamefont{S$\o$rensen}}
  \bibnamefont{and}
  \bibinfo{author}{\bibfnamefont{K.}~\bibnamefont{M$\o$lmer}},
  \bibinfo{journal}{Phys. Rev. Lett.} \textbf{\bibinfo{volume}{91}},
  \bibinfo{pages}{097905} (\bibinfo{year}{2003}).

\bibitem[{\citenamefont{Unanyan et~al.}(2002)\citenamefont{Unanyan,
  Fleischhauer, Vitanov, and Bergmann}}]{Unanyan2002}
\bibinfo{author}{\bibfnamefont{R.~G.} \bibnamefont{Unanyan}},
  \bibinfo{author}{\bibfnamefont{M.}~\bibnamefont{Fleischhauer}},
  \bibinfo{author}{\bibfnamefont{N.~V.} \bibnamefont{Vitanov}},
  \bibnamefont{and} \bibinfo{author}{\bibfnamefont{K.}~\bibnamefont{Bergmann}},
  \bibinfo{journal}{Phys. Rev. A} \textbf{\bibinfo{volume}{66}},
  \bibinfo{pages}{042101} (\bibinfo{year}{2002}).

\bibitem[{\citenamefont{Doherty et~al.}(2000)\citenamefont{Doherty, Habib,
  Jacobs, Mabuchi, and Tan}}]{Doherty2000}
\bibinfo{author}{\bibfnamefont{A.~C.} \bibnamefont{Doherty}},
  \bibinfo{author}{\bibfnamefont{S.}~\bibnamefont{Habib}},
  \bibinfo{author}{\bibfnamefont{K.}~\bibnamefont{Jacobs}},
  \bibinfo{author}{\bibfnamefont{H.}~\bibnamefont{Mabuchi}}, \bibnamefont{and}
  \bibinfo{author}{\bibfnamefont{S.~M.} \bibnamefont{Tan}},
  \bibinfo{journal}{Phys. Rev. A} \textbf{\bibinfo{volume}{62}},
  \bibinfo{pages}{012105} (\bibinfo{year}{2000}).

\bibitem[{\citenamefont{Belavkin}(1999)}]{Belavkin1999}
\bibinfo{author}{\bibfnamefont{V.}~\bibnamefont{Belavkin}},
  \bibinfo{journal}{Rep. on Math. Phys.} \textbf{\bibinfo{volume}{43}},
  \bibinfo{pages}{405} (\bibinfo{year}{1999}).

\bibitem[{\citenamefont{Geremia
  et~al.}(2003{\natexlab{b}})\citenamefont{Geremia, Stockton, Doherty, and
  Mabuchi}}]{Geremia2003a}
\bibinfo{author}{\bibfnamefont{J.~M.} \bibnamefont{Geremia}},
  \bibinfo{author}{\bibfnamefont{J.~K.} \bibnamefont{Stockton}},
  \bibinfo{author}{\bibfnamefont{A.~C.} \bibnamefont{Doherty}},
  \bibnamefont{and} \bibinfo{author}{\bibfnamefont{H.}~\bibnamefont{Mabuchi}},
  \bibinfo{journal}{Phys. Rev. Lett.} \textbf{\bibinfo{volume}{91}},
  \bibinfo{pages}{250801} (\bibinfo{year}{2003}{\natexlab{b}}).

\bibitem[{\citenamefont{Stockton
  et~al.}(2003{\natexlab{b}})\citenamefont{Stockton, Geremia, Doherty, and
  Mabuchi}}]{Stockton2003b}
\bibinfo{author}{\bibfnamefont{J.~K.} \bibnamefont{Stockton}},
  \bibinfo{author}{\bibfnamefont{J.}~\bibnamefont{Geremia}},
  \bibinfo{author}{\bibfnamefont{A.~C.} \bibnamefont{Doherty}},
  \bibnamefont{and} \bibinfo{author}{\bibfnamefont{H.}~\bibnamefont{Mabuchi}},
  \bibinfo{journal}{quant-ph/0309101}  (\bibinfo{year}{2003}{\natexlab{b}}).

\bibitem[{\citenamefont{Andre et~al.}(2004)\citenamefont{Andre, Sorensen, and
  Lukin}}]{Andre2004}
\bibinfo{author}{\bibfnamefont{A.}~\bibnamefont{Andre}},
  \bibinfo{author}{\bibfnamefont{A.~S.} \bibnamefont{Sorensen}},
  \bibnamefont{and} \bibinfo{author}{\bibfnamefont{M.~D.} \bibnamefont{Lukin}},
  \bibinfo{journal}{quant-ph/0401130}  (\bibinfo{year}{2004}).

\bibitem[{\citenamefont{Armen et~al.}(2002)\citenamefont{Armen, Au, Stockton,
  Doherty, and Mabuchi}}]{Armen2002}
\bibinfo{author}{\bibfnamefont{M.~A.} \bibnamefont{Armen}},
  \bibinfo{author}{\bibfnamefont{J.~K.} \bibnamefont{Au}},
  \bibinfo{author}{\bibfnamefont{J.~K.} \bibnamefont{Stockton}},
  \bibinfo{author}{\bibfnamefont{A.~C.} \bibnamefont{Doherty}},
  \bibnamefont{and} \bibinfo{author}{\bibfnamefont{H.}~\bibnamefont{Mabuchi}},
  \bibinfo{journal}{Phys. Rev. Lett} \textbf{\bibinfo{volume}{89}},
  \bibinfo{pages}{133602} (\bibinfo{year}{2002}).

\bibitem[{\citenamefont{Ahn et~al.}(2002)\citenamefont{Ahn, Doherty, and
  Landahl}}]{Ahn2002}
\bibinfo{author}{\bibfnamefont{C.}~\bibnamefont{Ahn}},
  \bibinfo{author}{\bibfnamefont{A.~C.} \bibnamefont{Doherty}},
  \bibnamefont{and} \bibinfo{author}{\bibfnamefont{A.~J.}
  \bibnamefont{Landahl}}, \bibinfo{journal}{Phys. Rev. A}
  \textbf{\bibinfo{volume}{65}}, \bibinfo{pages}{042301}
  (\bibinfo{year}{2002}).

\bibitem[{\citenamefont{Doherty and Jacobs}(1999)}]{Doherty1999}
\bibinfo{author}{\bibfnamefont{A.~C.} \bibnamefont{Doherty}} \bibnamefont{and}
  \bibinfo{author}{\bibfnamefont{K.}~\bibnamefont{Jacobs}},
  \bibinfo{journal}{Phys. Rev. A} \textbf{\bibinfo{volume}{60}},
  \bibinfo{pages}{2700} (\bibinfo{year}{1999}).

\bibitem[{\citenamefont{van Handel et~al.}(2004)\citenamefont{van Handel,
  Stockton, and Mabuchi}}]{vanHandel2004}
\bibinfo{author}{\bibfnamefont{R.}~\bibnamefont{van Handel}},
  \bibinfo{author}{\bibfnamefont{J.~K.} \bibnamefont{Stockton}},
  \bibnamefont{and} \bibinfo{author}{\bibfnamefont{H.}~\bibnamefont{Mabuchi}},
  \bibinfo{journal}{quant-ph/0402136}  (\bibinfo{year}{2004}).

\bibitem[{\citenamefont{Gardiner}(1985)}]{Gardiner2002}
\bibinfo{author}{\bibfnamefont{C.~W.} \bibnamefont{Gardiner}},
  \emph{\bibinfo{title}{Handbook of Stochastic Methods}}
  (\bibinfo{publisher}{Springer}, \bibinfo{address}{New York},
  \bibinfo{year}{1985}), \bibinfo{edition}{2nd} ed.

\bibitem[{\citenamefont{{\O}ksendal}(1998)}]{Oksendal1998}
\bibinfo{author}{\bibfnamefont{B.}~\bibnamefont{{\O}ksendal}},
  \emph{\bibinfo{title}{Stochastic Differential Equations}}
  (\bibinfo{publisher}{Springer Verlag}, \bibinfo{year}{1998}),
  \bibinfo{edition}{5th} ed.

\bibitem[{\citenamefont{Silberfarb and Deutsch}(2003)}]{Deutsch2003}
\bibinfo{author}{\bibfnamefont{A.}~\bibnamefont{Silberfarb}} \bibnamefont{and}
  \bibinfo{author}{\bibfnamefont{I.}~\bibnamefont{Deutsch}},
  \bibinfo{journal}{Phys. Rev. A} \textbf{\bibinfo{volume}{68}},
  \bibinfo{pages}{013817} (\bibinfo{year}{2003}).

\bibitem[{\citenamefont{Smith et~al.}(2003)\citenamefont{Smith, Chaudhury, and
  Jessen}}]{Jessen2003}
\bibinfo{author}{\bibfnamefont{G.~A.} \bibnamefont{Smith}},
  \bibinfo{author}{\bibfnamefont{S.}~\bibnamefont{Chaudhury}},
  \bibnamefont{and} \bibinfo{author}{\bibfnamefont{P.~S.}
  \bibnamefont{Jessen}}, \bibinfo{journal}{J. Opt. B: Quant. Semiclass. Opt.}
  \textbf{\bibinfo{volume}{5}}, \bibinfo{pages}{323} (\bibinfo{year}{2003}).

\bibitem[{\citenamefont{Stockton et~al.}(2002)\citenamefont{Stockton, Armen,
  and Mabuchi}}]{Stockton2002}
\bibinfo{author}{\bibfnamefont{J.}~\bibnamefont{Stockton}},
  \bibinfo{author}{\bibfnamefont{M.}~\bibnamefont{Armen}}, \bibnamefont{and}
  \bibinfo{author}{\bibfnamefont{H.}~\bibnamefont{Mabuchi}},
  \bibinfo{journal}{J. Opt. Soc. Am. B} \textbf{\bibinfo{volume}{19}},
  \bibinfo{pages}{3019} (\bibinfo{year}{2002}).

\bibitem[{\citenamefont{Bartlett and Wiseman}(2003)}]{Wiseman2003}
\bibinfo{author}{\bibfnamefont{S.~D.} \bibnamefont{Bartlett}} \bibnamefont{and}
  \bibinfo{author}{\bibfnamefont{H.~M.} \bibnamefont{Wiseman}},
  \bibinfo{journal}{Phys. Rev. Lett} \textbf{\bibinfo{volume}{91}},
  \bibinfo{pages}{097903} (\bibinfo{year}{2003}).

\bibitem[{\citenamefont{Cabello}(2002)}]{Cabello2002}
\bibinfo{author}{\bibfnamefont{A.}~\bibnamefont{Cabello}},
  \bibinfo{journal}{Phys. Rev. A} \textbf{\bibinfo{volume}{65}},
  \bibinfo{pages}{032108} (\bibinfo{year}{2002}).

\bibitem[{\citenamefont{Holstein and Primakoff}(1940)}]{Holstein1940}
\bibinfo{author}{\bibfnamefont{T.}~\bibnamefont{Holstein}} \bibnamefont{and}
  \bibinfo{author}{\bibfnamefont{H.}~\bibnamefont{Primakoff}},
  \bibinfo{journal}{Phys. Rev.} \textbf{\bibinfo{volume}{58}},
  \bibinfo{pages}{1098} (\bibinfo{year}{1940}).

\bibitem[{\citenamefont{Brody and Hughston}(2002)}]{Hughston2002}
\bibinfo{author}{\bibfnamefont{D.~C.} \bibnamefont{Brody}} \bibnamefont{and}
  \bibinfo{author}{\bibfnamefont{L.~P.} \bibnamefont{Hughston}},
  \bibinfo{journal}{quant-ph/0203035}  (\bibinfo{year}{2002}).

\bibitem[{\citenamefont{Wiseman et~al.}(2002)\citenamefont{Wiseman, Mancini,
  and Wang}}]{Wiseman2002}
\bibinfo{author}{\bibfnamefont{H.~M.} \bibnamefont{Wiseman}},
  \bibinfo{author}{\bibfnamefont{S.}~\bibnamefont{Mancini}}, \bibnamefont{and}
  \bibinfo{author}{\bibfnamefont{J.}~\bibnamefont{Wang}},
  \bibinfo{journal}{Phys. Rev. A} \textbf{\bibinfo{volume}{66}},
  \bibinfo{pages}{013807} (\bibinfo{year}{2002}).

\bibitem[{\citenamefont{Kloeden et~al.}(1997)\citenamefont{Kloeden, Platen, and
  Schurz}}]{Kloeden2003}
\bibinfo{author}{\bibfnamefont{P.~E.} \bibnamefont{Kloeden}},
  \bibinfo{author}{\bibfnamefont{E.}~\bibnamefont{Platen}}, \bibnamefont{and}
  \bibinfo{author}{\bibfnamefont{H.}~\bibnamefont{Schurz}},
  \emph{\bibinfo{title}{Numerical Solution of SDE Through Computer
  Experiments}} (\bibinfo{publisher}{Springer}, \bibinfo{address}{New York},
  \bibinfo{year}{1997}).

\end{thebibliography}

\end{document}